# Precision Measurement of the Weak Charge of the Proton

**The $Q_{weak}$ Collaboration:**

D. Androić[1], D.S. Armstrong[2], A. Asaturyan[3], T. Averett[2], J. Balewski[4], K. Bartlett[2], J. Beaufait[5], R.S. Beminiwattha[6], J. Benesch[5], F. Benmokhtar[7], J. Birchall[8], R.D. Carlini[5,2∗], J.C. Cornejo[2], S. Covrig Dusa[5], M.M. Dalton[9], C.A. Davis[10], W. Deconinck[2], J. Diefenbach[11], J.F. Dowd[2], J.A. Dunne[12], D. Dutta[12], W.S. Duvall[13], M. Elaasar[14], W.R. Falk[8,†], J.M. Finn[2,†], T. Forest[15,16], C. Gal[9], D. Gaskell[5], M.T.W. Gericke[8], J. Grames[17], V.M. Gray[2], K. Grimm[16,2], F. Guo[4], J.R. Hoskins[2], D. Jones[9], M. Jones[5], R. Jones[18], M. Kargiantoulakis[9], P.M. King[6], E. Korkmaz[19], S. Kowalski[4], J. Leacock[13], J. Leckey[2], A.R. Lee[13], J.H. Lee[6,2], L. Lee[10,8], S. MacEwan[8], D. Mack[5], J.A. Magee[2], R. Mahurin[8], J. Mammei[13], J.W. Martin[20], M.J. McHugh[21], D. Meekins[5], J. Mei[5], K.E. Mesick[21], R. Michaels[5], A. Micherdzinska[21], A. Mkrtchyan[3],H. Mkrtchyan[3], N. Morgan[13], A. Narayan[12], L.Z. Ndukum[12], V. Nelyubin[9], H. Nuhait[16], Nuruzzaman[11,12], W.T.H. van Oers[10,8], A.K. Opper[21], S.A. Page[8], J. Pan[8], K.D. Paschke[9], S.K. Phillips[22], M.L. Pitt[13], M. Poelker[17], J.F. Rajotte[4], W.D. Ramsay[10,8], J. Roche[6], B. Sawatzky[5], T. Seva[1], M.H. Shabestari[12], R. Silwal[9], N. Simicevic[16], G.R. Smith[5∗], P. Solvignon[5,†], D.T. Spayde[23], A. Subedi[12], R. Subedi[21], R. Suleiman[17], V. Tadevosyan[3], W.A. Tobias[9], V. Tvaskis[20,8], B. Waidyawansa[6], P. Wang[8], S.P. Wells[16], S.A. Wood[5], S. Yang[2], R.D. Young[24], P. Zang[25], and S. Zhamkochyan[3].    ∗Corresponding author, †Deceased

**Collaborating Institutions:**

[1]Department of Physics, University of Zagreb, Zagreb HR-10002, Croatia. [2]Department of Physics, William & Mary, Williamsburg, Virginia 23185, USA. [3]Division of Experimental Physics, A. I. Alikhanyan National Science Laboratory (Yerevan Physics Institute), Yerevan 0036, Armenia. [4]Department of Physics, Massachusetts Institute of Technology, Cambridge, Massachusetts 02139, USA. [5]Physics Division, Thomas Jefferson National Accelerator Facility, Newport News, Virginia 23606, USA. [6]Department of Physics and Astronomy, Ohio University, Athens, Ohio 45701, USA. [7]Department of Physics, Christopher Newport University, Newport News, Virginia 23606, USA. [8]Department of Physics and Astronomy, University of Manitoba, Winnipeg, Manitoba R3T 2N2, Canada. [9]Department of Physics, University of Virginia, Charlottesville, Virginia 22903, USA. [10]Science Division, TRIUMF, Vancouver, British Columbia V6T 2A3, Canada. [11]Department of Physics, Hampton University, Hampton, Virginia 23668, USA. [12]Department of Physics and Astronomy, Mississippi State University, Mississippi State, Mississippi 39762, USA. [13]Department of Physics, Virginia Polytechnic Institute and State University, Blacksburg, Virginia 24061, USA. [14]Department of Natural Sciences, Southern University at New Orleans, New Orleans, Louisiana 70126, USA. [15]Department of Physics, Idaho State University, Pocatello, Idaho 83209, USA. [16]Department of Physics, Louisiana Tech University, Ruston, Louisiana 71272, USA. [17]Accelerator Division, Thomas Jefferson National Accelerator Facility, Newport News, Virginia 23606, USA. [18]Department of Physics, University of Connecticut, Storrs-Mansfield, Connecticut 06269, USA. [19]Department of Physics, University of Northern British Columbia, Prince George, British Columbia V2N 4Z9, Canada. [20]Department of Physics, University of Winnipeg, Winnipeg, Manitoba R3B 2E9, Canada. [21]Department of Physics, George Washington University, Washington, D.C. 20052, USA. [22]Department of Physics, University of New Hampshire, Durham, New Hampshire 03824, USA. [23]Physics Department, Hendrix College, Conway, Arkansas 72032, USA. [24]Department of Physics and Mathematical Physics, University of Adelaide, Adelaide, South Australia 5005, Australia. [25]Department of Physics, Syracuse University, Syracuse, New York 13244, USA.

## Abstract:

The fields of nuclear and particle physics have undertaken extensive programs to search for evidence of physics beyond that explained by current theories. The observation of the Higgs boson at the Large Hadron Collider completed the set of particles predicted by the Standard Model (SM), currently the best description of fundamental particles and forces. However, the theory's limitations include a failure to predict fundamental parameters and the inability to account for dark matter/energy, gravity, and the matter-antimatter asymmetry in the universe, among other phenomena. Given the lack of additional particles found so far through direct searches in the post-Higgs era, indirect searches utilizing precise measurements of well-predicted SM observables allow highly-targeted alternative tests for physics beyond the SM. Indirect searches have the potential to reach mass/energy scales beyond those directly accessible by today's high-energy accelerators. The value of the weak charge of the proton $Q_W^p$ is an example of such an indirect search, as it sets the strength of the proton's interaction with other particles via the well-predicted neutral electroweak force. Parity symmetry (invariance under spatial inversion (x,y,z) → (-x,-y,-z)) is violated only in the weak interaction, thus providing a unique tool to isolate the weak interaction in order to measure the proton's weak charge[1]. Here we report $Q_W^p = 0.0719 \pm 0.0045$, as extracted from our measured parity-violating (PV) polarized electron-proton scattering asymmetry, $A_{ep} = -226.5 \pm 9.3$ ppb. Our value of $Q_W^p$ is in excellent agreement with the SM[2], and sets multi-TeV-scale constraints on any semi-leptonic PV physics not described within the SM.

In the electroweak SM, elastic scattering is mediated by the exchange of neutral currents (virtual photons and $Z^0$ bosons) between fundamental particles. A particle's weak charge $Q_W$ is analogous to but distinct from its electric charge $Q$: the former quantifies the vector coupling of the $Z^0$ boson to the particle, while the latter quantifies the vector coupling of the photon to the particle. The proton's weak charge $Q_W^p$ is defined[1] as the sum of the weak vector couplings $C_{1q}$ of the $Z^0$ boson to its constituent $u$ and $d$ quarks:

$$Q_W^p = -2(2C_{1u} + C_{1d}). \qquad (1)$$

The $Z^0$ exchange contribution to electron-proton scattering can be isolated via the weak interaction's unique PV signature (see Fig. 1). Interference between electromagnetic (EM) and weak scattering amplitudes leads to a PV asymmetry $A_{ep}$ that can be measured with a longitudinally-polarized electron beam incident on an unpolarized-proton target:

$$A_{ep} = \frac{\sigma_+ - \sigma_-}{\sigma_+ + \sigma_-} \qquad (2)$$

Here $\sigma_\pm$ represents the helicity-dependent elastic scattering $\vec{e}p$ cross section integrated over the scattered-electron detector acceptance. Helicity (±1) indicates the spin direction of the electrons in the beam as either parallel (+1) or anti-parallel (−1) to their momenta.

Measuring $A_{ep}$ at small four-momentum transfer ($Q^2$) suppresses contributions from the proton's extended structure relative to the proton's weak charge $Q_W^p$. However, $A_{ep}$ is smaller at smaller $Q^2$, making its measurement more challenging. In this low $Q^2$ limit, the PV asymmetry can be expressed[1] as

$$A_{ep}/A_0 = Q_W^p + Q^2 B(Q^2, \theta) \qquad (3)$$

where $A_0 = \frac{-G_F Q^2}{4\pi\alpha\sqrt{2}}$, $-Q^2$ is the four-momentum transfer squared, $B(Q^2, \theta)$ represents the proton's extended structure defined in terms of EM, strange and axial form factors, $\theta$ is the (polar) scattering angle of the electron in the lab frame with respect to the beam axis, $G_F$ is the Fermi constant, and $\alpha$ is the fine structure constant.

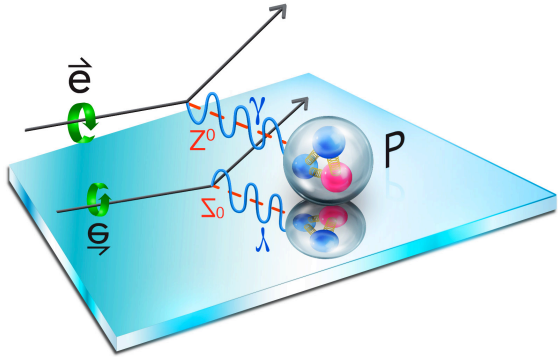

*Figure 1 | Parity-violating electron scattering from the proton.* The incoming electron with helicity +1 scatters away from the plane of the "Parity-Violating (PV) mirror" in this schematic. The image in the PV mirror shows the incoming electron with the opposite helicity −1, and instead of scattering into the plane of the PV mirror (as it would in a real mirror), it scatters out of the plane of the PV mirror. The dominant electromagnetic interaction, mediated by the photon ($\gamma$, blue wave), conserves parity. The weak interaction, mediated by the neutral $Z^0$ boson (dashed red line), violates parity. The weak interaction is studied experimentally by exploiting PV through reversals of the incident-beam helicity, which mimic the PV mirror "reflection".

The $Q_{weak}$ experiment[3,4] (see Extended Data Fig. 1) employed a beam of longitudinally-polarized electrons accelerated to 1.16 GeV at the Thomas Jefferson National Accelerator Facility. Three sequential acceptance-defining lead collimators matched to an eight-sector azimuthally-symmetric toroidal magnet selected electrons scattered from a liquid-hydrogen target at angles between 5.8° and 11.6°. In each magnet octant, elastically-scattered electrons were directed to a quartz detector fronted by lead pre-radiators. Cherenkov light produced by the EM shower passing through the quartz was converted to a current by photomultiplier tubes at each end of the quartz bars. These currents were integrated for each one-ms-long helicity state of the beam. For calibration purposes, and to demonstrate understanding of the acceptance and backgrounds, drift chambers were periodically inserted to track individual particles during dedicated periods of low-current running.

In order to achieve <10 ppb precision, this experiment pushed existing boundaries on many fronts: higher polarized-beam intensity (180 μA), faster beam-helicity reversal (960/s), better GeV-scale beam polarimetry[5] precision (±0.6%), higher liquid-hydrogen target[6] luminosity (1.7 x $10^{39}$/(cm²-s)) and cooling power (3 kW), and higher total detector rates (7 GHz). Following a brief commissioning period, the experiment was divided into two roughly six-month run periods, between which improvements were made primarily to polarimetry and helicity-correlated beam-monitoring and control instrumentation. Further details, including the backgrounds and corrections associated with each of the two halves of the experiment, are provided in Methods.

The asymmetry-measurement results are $A_{ep} = -223.5 \pm 15.0$ (stat) $\pm 10.1$ (syst) ppb in the first half, and $-227.2 \pm 8.3$ (stat) $\pm 5.6$ (syst) ppb in the second half of the experiment. They are in excellent agreement with each other, and consistent with our previously-published commissioning result[3]. Accounting for correlations in some systematic uncertainties between the two measurement periods, the combined result is $A_{ep} = -226.5 \pm 7.3$ (stat) $\pm 5.8$ (syst) ppb. The total uncertainty achieved (9.3 ppb) sets a new level of precision for parity-violating electron scattering (PVES) from a nucleus.

The relationship between the measured asymmetries $A_{ep}$ and the proton's weak charge $Q_W^p$ is expressed in equation (3), where the hadronic-structure-dependent B-term grows with momentum transfer $Q^2$. Higher $Q^2$ data from previous PVES experiments (see references in Methods) were included in a global fit[7,8,3] to constrain the proton structure contributions for the short extrapolation from our datum to $Q^2=0$ in order to determine $Q_W^p$, the intercept of equation (3). The average $Q^2$ of this experiment (0.0248 (GeV/c)$^2$) is much smaller than for any other PVES experiment used in this fit, with correspondingly smaller contributions from proton structure. The dominating precision of the $Q_{weak}$ measurement tightly constrains the fit near $Q^2=0$, where the connection to $Q_W^p$ can be made.

The parameters of the global fit[7,8,3] to the PVES data are the axial-electron, vector-quark weak coupling constants $C_{1u}$ and $C_{1d}$, the strange charge radius $\rho_s$ and strange magnetic moment $\mu_s$ (which characterize the strength of the proton's electric and magnetic strange-quark form factors), and the strength of the isovector axial form factor $G_A^{Z(T=1)}$. The EM form factors $G_E$ and $G_M$ used in the fit were taken from ref. 9; uncertainties in this input were accounted for in the result for $Q_W^p$ and its uncertainty.

The $\vec{e}p$ asymmetries shown in Fig. 2 were corrected[1,3] for the energy-dependent piece of the $\gamma Z$-box weak radiative correction[10–13] and its uncertainty. No other electroweak radiative corrections need to be applied to determine $Q_W^p$. However, ordinary EM radiative corrections (bremsstrahlung) were accounted for in the asymmetries used in the fit, including our datum. These and all other details of the fitting procedure, as well as a description of the corrections that went into the asymmetry for this experiment, are described in Methods.

The global fit is shown in Fig. 2 together with the $\vec{e}p$ data expressed as $A_{ep}(Q^2, \theta = 0)/A_0$. In order to isolate the $Q^2$ dependence for this figure, the $\theta$ dimension was projected to 0° by subtracting $[A_{calc}(Q^2, \theta) - A_{calc}(Q^2, \theta = 0)]$ from the measured asymmetries $A_{ep}(Q^2, \theta)$ as described in refs 8, 3. The fit includes all relevant $\vec{e}p$, $\vec{e}$-$^2$H, and $\vec{e}$-$^4$He PVES data (see Methods). The PVES database provides a data-driven (as opposed to a more theoretical) constraint on the nucleon structure uncertainties in the extrapolation to $Q^2=0$. We consider this the best method to provide our main result (denoted in Table 1 as "PVES fit"), which is $Q_W^p = 0.0719 \pm 0.0045$. We now discuss the sensitivity of this result to variations in the experimental and theoretical input used to determine it.

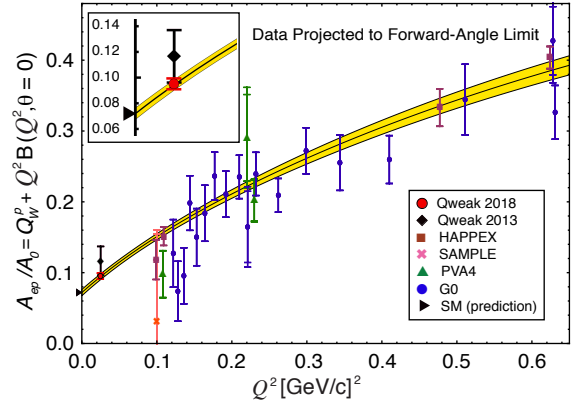

**Figure 2 The reduced asymmetry $A_{ep}/A_0 = Q_W^p + Q^2 B(Q^2, \theta = 0)$ vs $Q^2$.** The global fit is illustrated using $\vec{e}p$ asymmetries from this experiment, from the commissioning phase of this experiment[3] ($Q_{weak}$ 2013), as well as from earlier experiments HAPPEX, SAMPLE, PVA4, and G0 (see Methods for references), projected to $\theta = 0°$ and reduced by the factor $A_0(Q^2)$ appropriate for each datum. The data shown here include the $\gamma Z$-box radiative correction and uncertainty. Inner error bars (s.d.) include statistical and systematic uncertainties. Outer error bars on the data indicate the additional uncertainty estimated from the forward-angle projection. The solid line represents the global fit to the complete PVES database (see Methods), while the yellow band indicates the s.d. fit uncertainty. The arrowhead at $Q^2 = 0$ indicates the Standard Model (SM) prediction[2] for $Q_W^p = 0.0708(3)$, which agrees well with the intercept of the fit ($Q_W^p = 0.0719 \pm 0.0045$). The inset zooms in on the region around this experiment's result at $\langle Q^2 \rangle = 0.0248$ (GeV/c)$^2$.

| Method | Quantity | Value | Error |
|--------|----------|-------|-------|
| PVES fit | $Q_W^p$ | 0.0719 | 0.0045 |
| | $\rho_s$ | 0.20 | 0.11 |
| | $\mu_s$ | -0.19 | 0.14 |
| | $G_A^{Z(T=1)}$ | -0.64 | 0.30 |
| PVES fit + APV | $Q_W^p$ | 0.0718 | 0.0044 |
| | $Q_W^n$ | -0.9808 | 0.0063 |
| | $C_{1u}$ | -0.1874 | 0.0022 |
| | $C_{1d}$ | 0.3389 | 0.0025 |
| | $C_1$ correlation | -0.9318 | |
| PVES fit + LQCD | $Q_W^p$ | 0.0685 | 0.0038 |
| $Q_{weak}$ datum only | $Q_W^p$ | 0.0706 | 0.0047 |
| SM | $Q_W^p$ | 0.0708 | 0.0003 |

**Table 1: Results extracted from the asymmetry measured in the $Q_{weak}$ experiment.** The "PVES fit" method refers to a global fit incorporating the $Q_{weak}$ result and the PVES database described in Methods. When combined with APV[14,15] (to improve the $C_{1u,d}$ precision) this is indicated as method "PVES fit+APV". If the strange form factors in the global fit (without APV) are constrained to lattice QCD calculations[16], we label the result as "PVES + LQCD". The method labelled "$Q_{weak}$ datum only" uses the $Q_{weak}$ datum by itself, together with EM[9], strange[16], and axial[18] form factors from the literature in lieu of the global fit. Uncertainties are s.d.

Just as the proton's weak charge depends on its $u$ and $d$ quark content (see equation (1)), the weak charge of other nuclear systems depends on their (different) $u$ and $d$ quark content. Because $\vec{e}\text{-}{}^2\text{H}$, and $\vec{e}\text{-}{}^4\text{He}$ data are included in the global fit, $C_{1u}$ and $C_{1d}$ are each reasonably well determined. But if the very precise atomic parity-violation (APV) result[14,15] on ${}^{133}\text{Cs}$ is also included in the global fit, $C_{1u}$ and $C_{1d}$ can be determined with greater precision, and then used to extract the neutron's weak charge $Q_W^n = -2(C_{1u} + 2C_{1d})$. We note that inclusion or exclusion of the APV result has negligible impact on our result for $Q_W^p$, which is derived from the intercept of the global fit. The results for $C_{1u,d}$ and $Q_W^{p,n}$ obtained by including APV in the PVES global fit are denoted in Table 1 as method "PVES fit+APV". They agree with the SM values[2].

While our preferred result is based on the data-driven analysis of the "PVES fit" method, the final determination of the weak charge of the proton does not significantly change with additional theoretical constraint. One of the dominant uncertainties in the B-term of equation (3) arises from the knowledge of the strange-quark contributions. These have been determined very precisely in recent theoretical calculations[16,17] employing lattice quantum chromodynamics (LQCD). Using these theoretical results to constrain the extrapolation to $Q^2$=0 results in a slightly lower weak charge and a reduction in the uncertainty as shown in Table 1, method "PVES fit+LQCD". The APV result was not included in this determination of $Q_W^p$; its inclusion makes negligible difference.

Because the proximity to threshold ($Q^2 \to 0$) and precision of our $Q_{weak}$ result overwhelmingly dominate the fits described above, it is possible to go one step further and use the $Q_{weak}$ datum by itself to determine $Q_W^p$. The fact that the strange and axial form factors contribute so little at the kinematics of the $Q_{weak}$ experiment (0.1% and 2.5%, respectively) also helps motivate this consistency check. Using the same EM form factors[9] as in the fits above, the same lattice calculation[16] for the strange form factors, and following the extraction of ref. 18 for the axial form factor, the $Q_W^p$ result obtained using just the $Q_{weak}$ datum falls in-between the other consistent determinations described above which employ the entire PVES database (see Table 1, method "$Q_{weak}$ datum only"). The uncertainty of the $Q_W^p$ result using just the $Q_{weak}$ datum includes the additional uncertainty (4.6 ppb) due to the calculated form factors, but is only 4% larger than the global fit result uncertainty, which uses the entire PVES database. The dominant correction, from the EM form factors (23.7%), is well known in the low $Q^2$ regime of the $Q_{weak}$ experiment.

The determinations of $Q_W^p$ described above can be used to test the SM prediction of $\sin^2 \theta_W$, the fundamental electroweak parameter characterizing the mixing of the EM and weak interactions in the SM. Neglecting small quantum corrections, the SM predicts[19,20] $\sin^2 \theta_W$ in terms of the electroweak boson masses: $\sin^2 \theta_W = 1 - (M_W/M_Z)^2 \approx \frac{1}{4}$, and so $Q_W^p = 1 - 4\sin^2 \theta_W$ is nearly zero. This "accidental" SM suppression of $Q_W^p$ makes it an ideal observable to search for new PV interactions of natural size[21]. Using the latest input[2] to calculate quantum corrections which relate[1] $\sin^2 \theta_W$ $(Q = 0)$ to $Q_W^p$, as described in Methods, we obtain: $\sin^2 \theta_W$ $(Q=0)_{\overline{MS}} = 0.2383 \pm 0.0011$ in the modified-minimal-subtraction $(\overline{MS})$ scheme[2,19]. Subtracting 0.00012 in order to plot it at the $Q_{weak}$ energy scale ($Q=0.158$ GeV), our $\sin^2 \theta_W$ result is shown along with other determinations[2,20] in Fig. 3. It is consistent with the SM expectation and the purely leptonic E158 result[22] obtained in Møller ($\vec{e}e$) scattering, which has different sensitivities to new physics than our semi-leptonic ($\vec{e}p$) result.

Although the measurements at the $Z^0$-pole are more precise than our result, there exist a variety of beyond-the-Standard-Model (BSM) scenarios that can have significant influence on low-energy precision measurements while having little effect on collider measurements at the $Z^0$-mass energy scale[26]. A specific example is the dark-photon model of ref. 27, which allows large effects for few-hundred-MeV dark $Z$ mediators at low $Q$, but no effects at the $Z^0$ pole.

In order to explore this experiment's sensitivity to new BSM contact interactions, we follow the convention[28] where a "new physics" term $g^2/\Lambda^2$ is added to the SM term $g_{AV}^{eq}/(2v^2)$ in the Lagrangian for the neutral-current interaction of axial-vector electrons with vector quarks[‡]. Here

$g_{AV}^{eq} = C_{1q} = 2g_A^e g_V^q$ is the SM axial-electron, vector-quark coupling, $v^2 = \sqrt{2}/(2G_F)$, and $\Lambda$ represents the mass reach for new physics (the mass of the hypothetical BSM particle being exchanged) with coupling $g$. Expressed in terms of $Q_W^p$ and its uncertainty $\pm\Delta Q_W^p$, the 95% confidence level (CL) mass reach is

$$\frac{\Lambda_\pm}{g} = v\sqrt{\frac{4\sqrt{5}}{|Q_W^p \pm 1.96\Delta Q_W^p - Q_W^p(\text{SM})|}} \quad (4)$$

for which $\Lambda_+/g$ ($\Lambda_-/g$) is 7.4 (8.4) TeV.

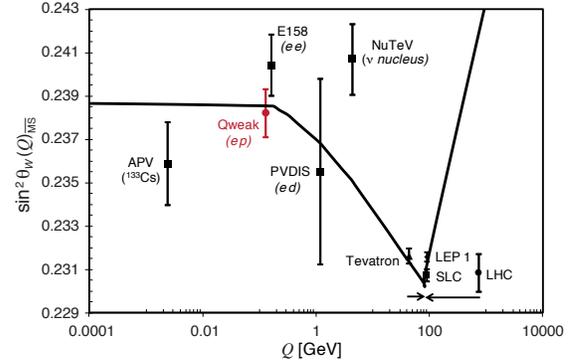

**Figure 3 | Variation of $\sin^2 \theta_W$ with energy scale $Q$.** The modified-minimal-subtraction $(\overline{MS})$ scheme is shown as the solid curve[19,2], together with experimental determinations from ref. 2 at the $Z^0$-pole[2] (Tevatron, LEP1, SLC, LHC), atomic parity violation on Cesium (APV)[14,15], Møller scattering (E158)[22], deep inelastic $\vec{e}$-$^2$H scattering (PVDIS)[23], and from neutrino-nucleus scattering (NuTeV)[24]. It has been argued[25], however, that the latter result contains significant unaccounted-for nuclear-physics effects. Our new result is plotted in red at the energy-scale of the $Q_{weak}$ experiment, $Q=0.158$ GeV (slightly offset horizontally for clarity). Error bars (s.d.) include statistical and systematic uncertainties.

For the extreme contact-interaction coupling[29] $g^2 = 4\pi$, the maximum mass reach for new semi-leptonic physics determined by our $Q_W^p$ result is $\Lambda_+ = 26.3$ TeV. Using the coupling $g^2 = 4\pi\alpha$ typically assumed for leptoquarks[30] (hypothetical BSM particles with both lepton and baryon

---

[‡] This convention[28] differs from an earlier one[1] by a factor of 4 (2 in $\Lambda$).

numbers), equation (4) rules out leptoquark masses below 2.3 TeV.

In a more general case[8] where the coupling of the new physics to the quarks is not restricted to the same 2:1 ratio of $u$ to $d$ quarks as the proton, the mass reach $\Lambda/g$ can be expressed as a circle of radius $(g/\Lambda)^2 \sqrt{2}/G_F$, about the SM origin $(C_{1u}^{SM}, C_{1d}^{SM}) = (-0.1887, 0.3419)$ in $C_{1q}$ space. This is illustrated in Fig. 4a, where the complementary constraints on the $C_{1q}$ provided by the weak charge measurements of this experiment ($Q_W^p = -2(2C_{1u} + C_{1d})$) and the $^{133}$Cs APV[14,15] result[2] ($Q_W^{Cs} = -2\{55(2C_{1u} + C_{1d}) + 78(C_{1u} + 2C_{1d})\}$) are also shown. Expressing the isospin dependence of potential BSM physics in terms of $h_v^u = \cos\theta_h$ and $h_v^d = \sin\theta_h$ as in ref. 8, 95% CL bounds are shown in Fig. 4b from this experiment, from APV[14,15,2], and from combining those two constraints. New physics is ruled out below the curves. Our combined constraint raises the $\theta_h$-independent limit for generic new semi-leptonic PV BSM physics to 3.5 TeV.

The low-energy precision frontier continues to offer an exciting landscape to search for BSM physics. The results of this experiment are consistent with the SM, and place important limits on new BSM physics. Future experiments propose to provide even more precise (and much more challenging) determinations of $Q_W^p$ at lower $Q^2$ (ref. 31), and of $Q_W^e$ (ref. 32), by taking the techniques and lessons learned in this experiment to the next level.

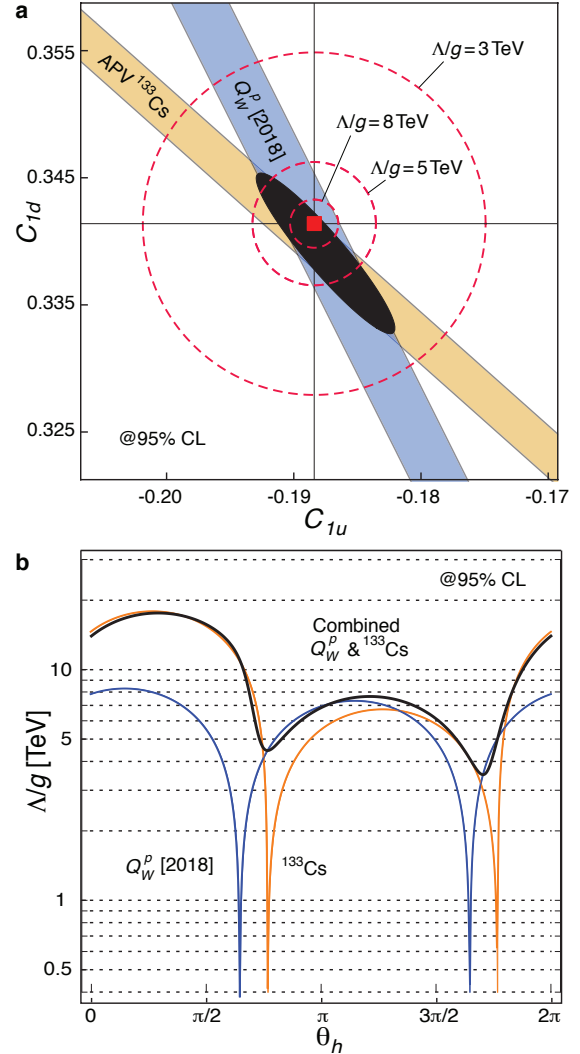

*Figure 4* | **Mass/coupling constraints on New Physics. (a)** 95% confidence-level (CL) constraints on the axial-electron, vector-quark weak coupling constants $C_{1u}$ and $C_{1d}$ provided by the weak charge determined in this experiment using the global fit method "PVES fit" (blue band) and the atomic parity violation result[14,15] on $^{133}$Cs (gold band). The combined (95% CL) constraint is shown by the black ellipse. Contours of the mass reach $\Lambda/g$ for new physics with coupling $g$ to arbitrary quark flavor ratios are indicated by dashed circles centered about the Standard Model values[2] of $C_{1u}$ and $C_{1d}$ denoted by the red square. **(b)** Mass reach $\Lambda/g$ (95% CL) as a function of the quark flavor-mixing angle $\theta_h$ for the $Q_{weak}$ "PVES fit" result (blue curve), for the $^{133}$Cs APV[14,15] result[2] (gold curve), and for both results combined (black curve). The two maxima in the blue curve at $\theta_h = \tan^{-1}(n_d/n_u) = \tan^{-1}(1/2) = 26.6°$ and 206.6° correspond to $\Lambda_-/g = 8.4$ TeV and $\Lambda_+/g = 7.4$ TeV in equation (4), respectively.

## Online Content

Methods, along with additional Extended Data display items and Source Data, are available in the online version of the paper; references unique to these sections appear only in the online paper.

## Acknowledgements

This work was supported by DOE Contract No. DEAC05-06OR23177, under which Jefferson Science Associates, LLC operates Thomas Jefferson National Accelerator Facility. Construction and operating funding for the experiment was provided through the U.S. Department of Energy (DOE), the Natural Sciences and Engineering Research Council of Canada (NSERC), the Canadian Foundation for Innovation (CFI), and the National Science Foundation (NSF) with university matching contributions from William and Mary, Virginia Tech, George Washington University, and Louisiana Tech University. We thank the staff of Jefferson Lab, in particular the accelerator operations staff, the target and cryogenic groups, the radiation control staff, as well as the Hall C technical staff for their help and support. We are grateful for the contributions of our undergraduate students. We thank TRIUMF for its contributions to the development of the spectrometer and integrating electronics, and BATES for its contributions to the spectrometer and Compton polarimeter. We are indebted to P.G. Blunden, J.D. Bowman, J. Erler, N.L. Hall, W. Melnitchouk, M.J. Ramsey-Musolf, and A.W. Thomas for many useful discussions. We also thank P.A. Souder for contributions to the experiment's analysis. Figure 2 adapted with permission from Ref. 3. Copyrighted by the American Physical Society.

## Author Contributions

Authors contributed to one or more of the following areas: proposing, leading, or running the experiment; design, construction, optimization, and testing of the experimental apparatus or data acquisition system; data analysis; simulation; extraction of the physics results from measured asymmetries; and the writing of this Letter.

## Author Information

Reprints and permissions information is available at www.nature.com/reprints. The authors declare no competing financial interests. Readers are welcome to comment on the online version of the paper. Correspondence and requests for materials should be addressed to R.D.C. (carlini@jlab.org) or G.R.S. (smithg@jlab.org).

## METHODS

Here we describe the formalism connecting the experimentally measured asymmetry $A_{ep}$ to the proton's weak charge $Q_W^p$, the experimental method used to determine $A_{ep}$, some measures of the data quality, and the electroweak radiative corrections needed to extract the weak mixing angle from $Q_W^p$.

**Formalism.** The $Q_{weak}$ experiment measured the parity-violating asymmetry $A_{ep}$, which is the normalized difference between the cross section ($\sigma$) of elastic scattering of longitudinally polarized electrons with positive ($\sigma_+$) and negative ($\sigma_-$) helicity from unpolarized protons:

$$A_{ep} = \frac{\sigma_+ - \sigma_-}{\sigma_+ + \sigma_-} \qquad (5)$$

The extended structure of the proton can be represented using form factors. Assuming charge symmetry, the single boson exchange ("tree-level") expression for the asymmetry can be written in terms of the proton and neutron electromagnetic form factors $G_E^p$, $G_M^p$, $G_E^n$, and $G_M^n$, the strange electric and magnetic form factors $G_E^s$ and $G_M^s$, and the neutral weak axial form factor $G_A^p$ as[7,33]

$$A_{ep} = A_0 \, \frac{A_V^p + A_S^p + A_A^p}{\varepsilon \left(G_E^p\right)^2 + \tau \left(G_M^p\right)^2}, \qquad (6)$$

with

$A_0 = \left[\frac{-G_F Q^2}{4\pi\alpha\sqrt{2}}\right]$,

$A_V^p = Q_W^p \left[\varepsilon \left(G_E^p\right)^2 + \tau \left(G_M^p\right)^2\right] - \left[\varepsilon G_E^p G_E^n + \tau G_M^p G_M^n\right]$,

$A_S^p = -\varepsilon G_E^p G_E^s - \tau G_M^p G_M^s$,

$A_A^p = -(1 - 4\sin^2\theta_W)\varepsilon' G_M^p G_A^p$,

$Q_W^p = -2(2C_{1u} + C_{1d})$,

where

$$\epsilon = \frac{1}{1 + 2(1+\tau)\tan^2\frac{\theta}{2}}, \varepsilon' = \sqrt{\tau(1+\tau)(1-\varepsilon^2)}$$

are kinematic quantities, $\alpha$ the fine structure constant, $G_F$ the Fermi constant, $\theta_W$ the weak mixing angle, $-Q^2$ the four-momentum transfer squared, $\theta$ is the laboratory electron scattering angle, and $\tau = Q^2/4M^2$, where $M$ is the proton mass.

Taking the limits $\theta \to 0$ and $Q^2 \to 0$, equation (6) reduces[34] to equation (3). Similar relationships apply for parity-violating elastic scattering from $^4$He and quasielastic scattering from $^2$H, but involve different linear combinations of $C_{1u}$ and $C_{1d}$ and the various form factors[33,35].

**Global Fit.** A global fit was made of equation (6) to a database which included all relevant PVES data up to $Q^2 = 0.63$ (GeV/c)$^2$: 28 proton results from the G0[36,37], HAPPEX[38-41], SAMPLE[42], PVA4[43-45] and $Q_{weak}$[3] collaborations, including the present result, two $^4$He elastic results (HAPPEX[40,46]) and five $^2$H quasi-elastic results (G0[37], PVA4[45,47] and SAMPLE[48]).

The fit followed the procedure introduced by Young et al.[7], and as used in refs 3 and 8. There were six parameters in the fit: $C_{1u}$, $C_{1u}$, and four that characterize the strange and axial form factors: the strange radius $\rho_s$, the strange magnetic moment $\mu_s$, and the magnitudes of $G_A^p$ and $G_A^n$. The parameterizations chosen for the strange form factors were $G_E^s = \rho_s Q^2 G_D$ and $G_M^s = \mu_s G_D$, with the dipole form $G_D = (1 - Q^2/\lambda^2)^{-2}$ where $\lambda = 1$ GeV/c. Young et al.[7] examined the consequences of using more elaborate $Q^2$ dependences for $G_E^s$ and $G_M^s$ and found that the data do not require their adoption. More recent analyses have come to the same conclusion[18,49]. Lattice QCD calculations[16,17] have also found similar shapes for $G_E^s$ and $G_M^s$. The neutron's axial form factor $G_A^n$ enters the fit through the inclusion of $^2$H data in the database. For the isovector combination $G_A^{Z(T=1)} = \left(G_A^p - G_A^n\right)/2$, the dipole form $G_D$ was again adopted, with the normalization as the parameter. The isoscalar combination $G_A^{Z(T=0)} = \left(G_A^p + G_A^n\right)/2$ is known to be small theoretically, so it was constrained in the fit to the theoretical value[50] $G_A^{Z(T=0)} = -0.08 \pm 0.26$, reducing the effective number of parameters to five. The unconstrained isovector combination $G_A^{Z(T=1)}$ presented in Table 1 was constructed from the values of $G_A^p$ ($G_A^n$) determined in the fit, which were $-0.59 \pm 0.34$ ($0.68 \pm 0.44$) with a covariance of $-0.0265$.

The electromagnetic form factors $G_{E,M}^{p,n}$ were taken from ref. 9. If, instead, we use any one of several alternate parameterizations of these form factors[51-53], the fitted result for $Q_W^p$ changes by less than 1%. We have incorporated this range as a systematic uncertainty.

Each of the experimental asymmetries $A_{ep}$ used in the fit needed to be corrected for the one electroweak radiative correction with significant energy dependence, the $\gamma Z$ box diagram[1]. Three independent theoretical groups have calculated this correction, with results in excellent mutual agreement[10,13,54]. We have adopted the most recent, data-driven calculation of the vector[10] (0.0054(4) and axial-vector[11,12] ($-0.0007(2)$) contributions, multiplied by a small $Q^2$ correction[13] (0.978(12)); the numerical values here are in terms of the effect on the reduced asymmetry $A_{ep}/A_0$. The total correction to our datum is 0.0046(5), corresponding to a 6.4% $\pm$ 0.6% correction to $Q_W^p$. If we choose instead to use either the calculations of ref. 13 or ref. 54, the extracted value of $Q_W^p$ is essentially unchanged, and the effect on its uncertainty is negligible.

The fit was done using linear $\chi^2$ minimization, with a resulting $\chi^2/\text{dof} = 1.25$ for 29 degrees of freedom (dof).

**Apparatus.** The $Q_{\text{weak}}$ experiment was performed with a custom apparatus installed in Hall C at the Thomas Jefferson National Accelerator Facility. A complete description of the apparatus and critical aspects of the accelerator can be found in ref. 4. Here we highlight some of the most important details illustrated in Extended Data Fig.1.

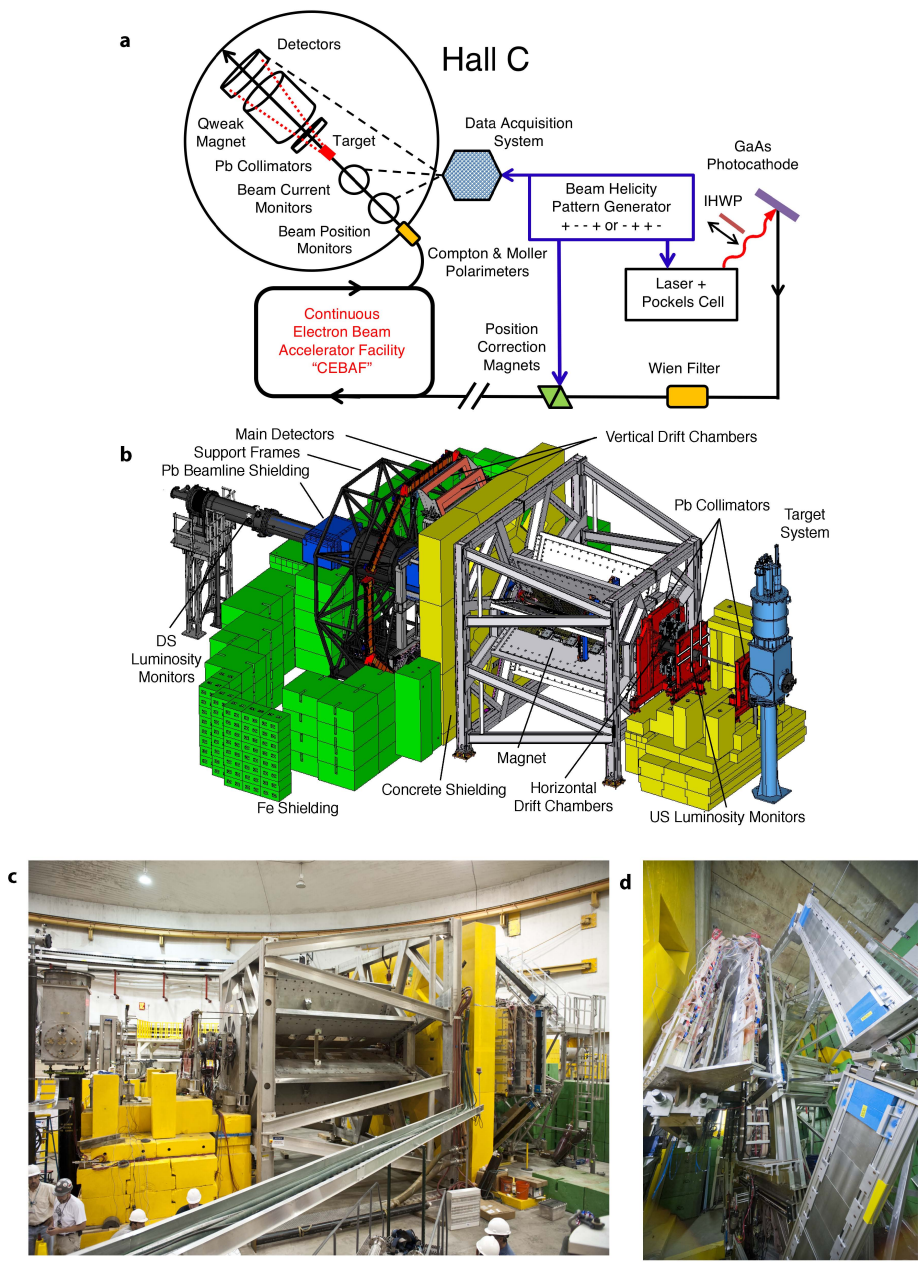

**Extended Data Figure 1 | Apparatus.** (**a**) Schematic of critical accelerator components and the $Q_{weak}$ apparatus[4]. The electron beam is generated at the photocathode, accelerated by the Continuous Electron Beam Accelerator Facility (CEBAF), and sent to experimental Hall C, where it is monitored by Beam Position Monitors (BPMs) and Beam Current Monitors (BCMs). The insertable half-wave plate (IHWP) provides a slow reversal of the electron beam helicity. The data acquisition system (DAQ) records the data. (**b**) CAD view of the experimental apparatus. (**c**) $Q_{weak}$ apparatus before final shielding configuration was installed. (**d**) Interior of shielded detector hut showing two of the Cherenkov detectors (right) and a pair of tracking chambers (left).

The electron beam was longitudinally polarized and its helicity was reversed at a rate of 960 Hz in a pseudorandom sequence of "helicity quartets" $(+ - - +)$ or $(- + + -)$. The quartet pattern minimized noise due to slow linear drifts, while the rapid helicity reversal limited noise due to fluctuations in the target density and beam properties. As a test for possible false asymmetries, two methods were used to reverse the beam helicity on a slower timescale than the rapid helicity reversal. About every 8 hours, the helicity of the laser beam used to generate the polarized electron beam was reversed by insertion of a half-wave plate (IHWP) in its path. Monthly, the helicity of the electron beam in the accelerator's injector region was reversed using a "double Wien" spin rotator[55]. Non-vanishing correlations between the properties of the electron beam (intensity, position, angle, energy) and its helicity lead to false contributions to the measured asymmetry. These helicity-correlated beam properties were suppressed by careful setup of the helicity-defining optics and active feedback systems in the polarized injector laser system. The electron beam transport line was instrumented to allow for correction of residual helicity-correlated beam properties. Beam monitors upstream of the experimental apparatus provided continuous, non-invasive measurement of the electron beam's intensity, position, angle, and energy. The response of the experimental apparatus to changes in the beam properties was periodically measured using a beam modulation system that generated controlled variations in the beam's position and angle with magnets and in its energy with a radio-frequency accelerating cavity. Finally, the electron beam's polarization was measured with two independent polarimeters upstream of the $Q_{weak}$ apparatus.

The experiment routinely ran with $180 \, \mu A$ of 1.16 GeV, $\approx 88\%$ longitudinally polarized electrons incident on a 34.4 cm, 20 K liquid hydrogen target in an aluminum cell. A set of three lead collimators restricted the polar scattering angle acceptance to the range $5.8° \leq \theta \leq 11.6°$ with an azimuthal angle coverage of 49% of $2\pi$. A resistive toroidal magnet between the target and detectors separated the elastically scattered electrons of interest from irreducible backgrounds arising from inelastic and Møller scattering. The resulting scattered electrons were detected in eight synthetic quartz Cherenkov detectors, each 200 cm × 18 cm × 1.25 cm, arranged in an azimuthally symmetric pattern about the beam axis. The four-fold azimuthal symmetry was important to minimize and characterize the effects of helicity-correlated beam properties and residual transverse beam polarization. These "main detectors" were equipped with 2 cm thick Pb pre-radiators that amplified the electron signal and suppressed low-energy backgrounds. Light was collected from the detectors by photomultiplier tubes (PMT) at each end. The high rates of $\approx 880$ MHz per detector required a current-mode readout, in which the PMT anode current was converted to a voltage that was digitized and integrated every $\approx$millisecond. Photographs of the apparatus are in Extended Data Fig. 1.

The collimator/magnet configuration and carefully designed shielding were effective in minimizing reducible sources of background such as direct line-of-sight (neutral) tracks originating in the target and secondary scattering from the beampipe. The residual diffuse background was monitored by background detectors in the main-detector shield-house. They consisted of a complete main detector placed in the super-elastic region, a dark box with a bare photomultiplier tube, and a dark box with a photomultiplier tube attached to a light guide. A symmetric array of four smaller detectors (the "upstream luminosity monitors") placed on the upstream face of the defining (middle) collimator was effective in monitoring residual backgrounds from the tungsten beam collimator that shielded the downstream region from small angle scattered particles.

Finally, a tracking system consisting of drift chambers located before and after the magnet was deployed periodically to verify the acceptance-weighted kinematic distribution and to help study backgrounds. These measurements were done in special run periods at very low beam currents (0.1 - 200 nA) and using conventional individual pulse counting.

Data from the experiment's short commissioning run (4% of the size of the data set reported here) have been previously published[3]. Here we report the combined result from two run periods (referred to as Run 1 and 2), each about six months in duration. Improvements to the apparatus were made between the running periods and beam conditions were different in the two run periods, so we report below the values of the corrections and systematic uncertainties for each run period separately. To prevent possible biases in the analysis, the main-detector asymmetries were blinded by an additive shift in the asymmetry, different for each run period. When the data analysis was complete, the asymmetries were unblinded, revealing the results presented here.

**Data Analysis.** The raw asymmetry $A_{raw}$ was formed from the difference over the sum of the beam-charge normalized detector PMT signals (see equation (1)), summed over the 8 detectors. The measured asymmetry $A_{msr}$ was calculated from $A_{raw}$ by correcting for a variety of effects that could cause false asymmetries:

$$A_{msr} = A_{raw} + A_T + A_L + A_{BCM} + A_{BB} + A_{beam} + A_{bias}. \tag{7}$$

The methods for determining each of these corrections and their uncertainties are discussed below.

**$A_T$:** A small ($\approx 2\%$) residual transverse component of the incident beam polarization resulted in a transverse asymmetry $A_T$ (which is driven by a parity-conserving two-photon-exchange amplitude[56]). We determined this correction by measuring the transverse asymmetry using a maximally transversely-polarized incident beam[50]. An upper bound on the broken symmetry of the spectrometer/detector system (< 1.3%), was determined using this transversely polarized beam. Combining these effects with the measured residual transverse polarization components of the beam during data-taking with nominally longitudinally-polarized beam, we determined the correction to be $A_T = 0.0 \pm 1.1$ ppb and $0.0 \pm 0.7$ ppb for Run 1 and 2, respectively.

**$A_L$:** The non-linearity of the Cherenkov detector readout chain (photomultiplier tube, low noise voltage-to-current preamplifier, and analog-to-digital converter) was directly measured using a system with light-emitting diodes to be $0.7 \pm 0.5\%$ at the signal levels corresponding to those experienced during the experiment. This results in an additive correction $A_L$ of $1.3 \pm 1.0$ ppb and $1.2 \pm 0.9$ ppb for Run 1 and Run 2, respectively.

**$A_{BCM}$:** The beam current was measured non-invasively with radio-frequency resonant cavities to allow a precise relative comparison of the beam charge in each helicity state. Two such beam current monitors (BCM) were used in the Run 1 analysis, while three BCMs were used in the Run 2 analysis after the installation of an additional monitor and improvement of the low-noise digital demodulation electronics. The correction $A_{BCM}$ is zero by definition, since we normalize the integrated detector signals to the average of the BCMs in each period. The systematic uncertainty on this correction is determined from the variation in the reported charge asymmetry for the BCMs in each analysis, resulting in $\delta A_{BCM} = \pm 4.4$ ppb and $\delta A_{BCM} = \pm 2.1$ ppb for Run 1 and 2, respectively.

**$A_{BB}$:** A false asymmetry, caused by secondary events scattered from the beamline and the tungsten beam collimator, is referred to as the beamline-background asymmetry $A_{BB}$. These events were determined to be predominantly due to low-energy neutral particles. They contributed a small amount to the signal (0.19%), but they had a large asymmetry, believed to be associated with a helicity-dependent intensity or position variation in the extended halo around the main accelerated beam. While only a small component of the main detector signal, it dominated the asymmetry measurement of the background detectors, which were highly correlated (see

Extended Data Fig. 2b). A direct correlation between the main-detector asymmetry from these events and the background asymmetries measured by the background detectors was shown by blocking two of the eight openings in the first of the three Pb collimators with 5.1 cm thick tungsten plates (see Extended Data Fig. 2a).

To correct for this false asymmetry, a correlation factor was extracted (separately for Run 1 and 2) between the asymmetries of the main detector array and the upstream luminosity monitor array, as shown in Extended Data Fig. 2c. The correlation factor was combined with the measured upstream luminosity monitor asymmetry, averaged every 6 minutes, to correct the main-detector asymmetry in that interval. The resulting net corrections for Run 1 and 2, respectively, were $A_{BB} = 3.9 \pm 4.5$ ppb and  -2.4 $\pm$ 1.1 ppb, where the uncertainty includes contributions from the statistical error on the determination of the correlation and systematic errors extracted by allowing the correlation factor to vary randomly within a reasonable range over different time periods[58].

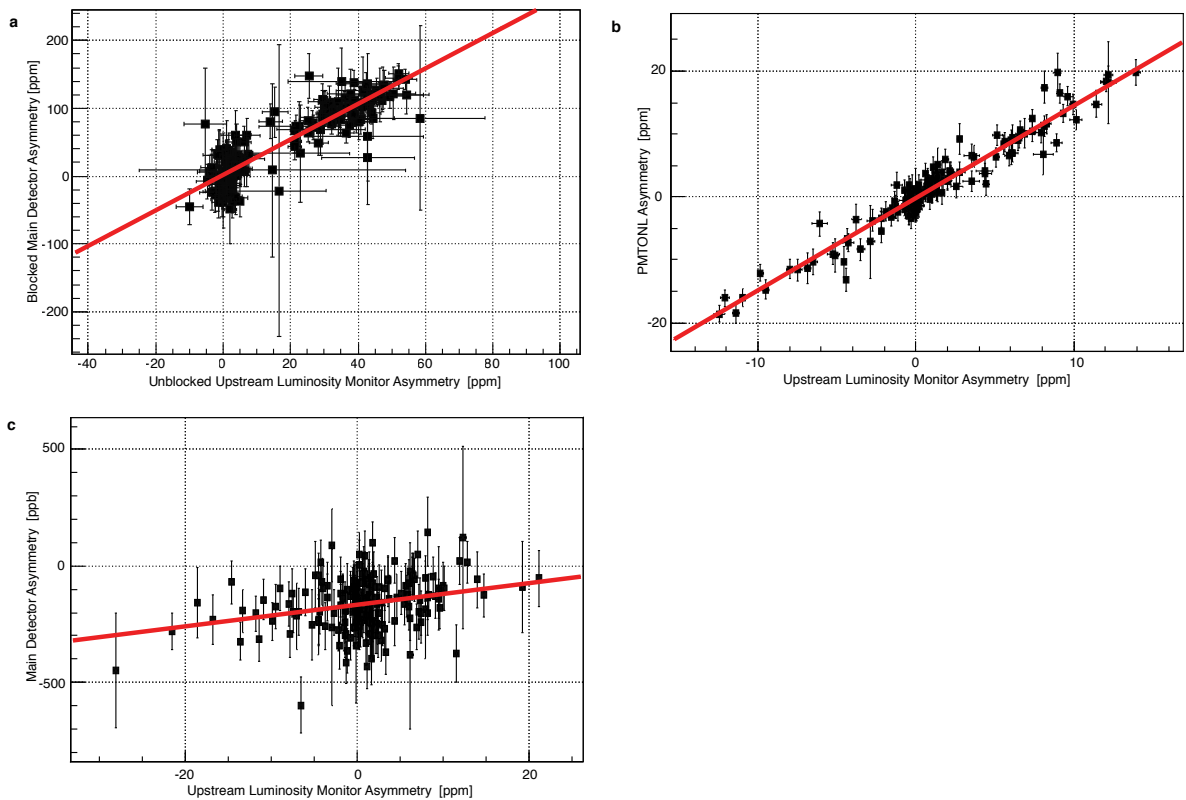

**Extended Data Figure 2 | Beamline background.** Determination of $A_{BB}$, the false asymmetry arising from beamline background events. Uncertainties are s.d. (**a**) Correlation of the main detector asymmetry to that of the upstream luminosity monitors, taken when the elastic electron signal for the main detectors was blocked at the first collimator. (**b**) Correlation of asymmetries from the upstream luminosity monitors with one of the other background detectors (PMTONL, which was a bare photomultiplier tube located in the detector shield house). (**c**) Correlation of the unblocked main detector asymmetry to that of the upstream luminosity monitor, for Run2. Our $A_{BB}$ determination was based on this slope.

**$A_\mathbf{beam}$:** Residual non-vanishing correlations in the properties of the electron beam were accounted for through a correction $A_\mathbf{beam}$. Five beam properties – transverse beam position and angle (horizontal: $X$, $X'$, vertical: $Y$, $Y'$), as well as energy – were monitored continuously as described above. The run-averaged helicity-correlated values of these parameters are presented in Extended Data Table 1. Using the measured helicity-correlated beam differences $\Delta\chi_i$ and the sensitivity of the measured asymmetry to variations in the beam parameters $\partial A/\partial\chi_i$, the corrections for the $i^{th}$ beam parameter were combined into an overall correction: $A_\mathbf{beam} = -\sum_{i=1}^{5}(\frac{\partial A}{\partial\chi_i})\Delta\chi_i$ .

The detector sensitivities $\partial A/\partial\chi_i$ were measured routinely by deliberately varying the beam parameters using the beam modulation system described above. This system made small perturbations of the beam parameters about their nominal values with a sinusoidal driving function at 125 Hz. Furthermore, an accelerator fast feedback system, that had 125 Hz in its operational frequency range was used, resulting in the beam position and angle parameters also being driven with a sinusoidal pattern 90° out of phase with respect to the beam modulation system. The result was that the beam parameters were driven in more than one way, allowing for redundant measurements of the detector sensitivities by using different combinations of the driven signals. The spread in those results was the dominant component of the systematic uncertainty for this correction. Typical values of the sensitivities are shown in Extended Data Table 1. For a perfectly symmetric apparatus, the position and angle sensitivities would be zero. For this apparatus, the sensitivities show that the horizontal plane had a larger broken symmetry than the vertical plane. The resulting corrections for Runs 1 and 2, respectively, were $A_\mathbf{beam}$ = 18.5 ± 4.1 ppb and 0.0 ± 1.1 ppb. The significantly smaller correction in Run 2 was due to smaller position and angle helicity-correlated differences during that period.

| Beam Parameter | Run 1 $\Delta\chi_i$ | Run 2 $\Delta\chi_i$ | Typical $\partial A/\partial\chi_i$ |
|---|---|---|---|
| $X$ | $-3.5 \pm 0.1$ nm | $-2.3 \pm 0.1$ nm | $-2$ ppb/nm |
| $X'$ | $-0.30 \pm 0.01$ nrad | $-0.07 \pm 0.01$ nrad | 50 ppb/nrad |
| $Y$ | $-7.5 \pm 0.1$ nm | $0.8 \pm 0.1$ nm | $< 0.2$ ppb/nm |
| $Y'$ | $-0.07 \pm 0.01$ nrad | $-0.04 \pm 0.01$ nrad | $< 3$ ppb/nrad |
| Energy | $-1.69 \pm 0.01$ ppb | $-0.12 \pm 0.01$ ppb | $-6$ ppb/ppb |

**Extended Data Table 1 | Helicity-correlated beam parameter differences and sensitivities.** The beam parameter differences and typical detector sensitivities for the five measured beam parameters, for both Run 1 and Run 2. Uncertainties are s.d.

**$A_\mathbf{bias}$:** When comparing the measured asymmetry in the two photomultiplier tubes, "left" and "right", which read out the signal at each end of each of the eight main detectors, a consistent difference of ≈ +300 ppb was found, as shown in Extended Fig. 3a. Here, "right" is beam direction $\hat{\boldsymbol{k}}$ crossed with radial direction $\hat{\boldsymbol{r}}$, $\hat{\boldsymbol{k}} \times \hat{\boldsymbol{r}}$. The effect is due to a left/right analyzing power in the multiple scattering of transversely (radially) polarized electrons through the lead pre-radiators of the main detectors. For perfect symmetry, this parity-conserving effect cancels when forming the parity-violating asymmetry of interest. Properly accounting for the minor broken symmetries of the as-built apparatus leads to a correction, $A_\mathbf{bias}$, referred to as rescattering bias.

A schematic of the physical model for this effect is shown in Extended Data Fig. 3a. Scattered electrons, which are initially fully longitudinally polarized, acquire some transverse polarization through precession as they transport through the spectrometer's magnetic field.

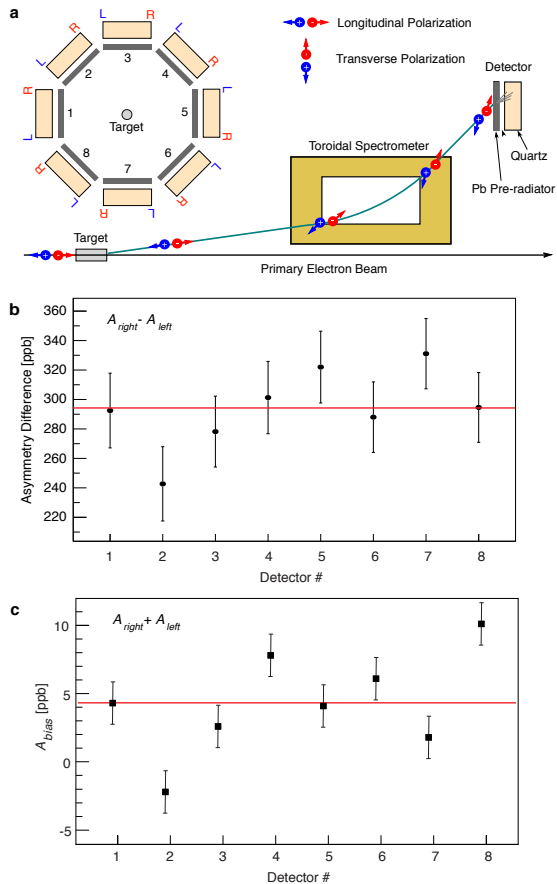

**Extended Data Figure 3 | Rescattering bias.** (**a**) Schematic illustrating the precession of longitudinally polarized electrons through the spectrometer magnet, generating significant transverse spin components upon arrival at the detector array (spin directions indicated by red and blue arrows for the two electron helicity states); end view of the detector array indicating "right" (R) and "left" (L) photomultiplier tube (PMT) positions shown on the left. (**b**) Difference between the asymmetry measured by the two PMT tubes "right" and "left" from each detector plotted vs. detector number (Run 2 data). (**c**) Calculated rescattering bias $A_{bias}$ vs. detector number, with the 8-detector averaged value shown by the red lines. Uncertainties (s.d.) are systematic.

The effect was modeled using a detailed Geant4[59] simulation of the transport of the detected electrons through the spectrometer and the lead radiator, including electromagnetic showering and Mott scattering. An asymmetry in the distribution of electrons penetrating the radiator develops due to the analyzing power of low-energy Mott scattering. A possible analyzing power for high-energy scattering due to non-Born processes[60] was also considered, but reasonable models for those processes[61] showed an insignificant contribution.

Asymmetries from this simulation (and from a variety of analytic effective models that reproduced the key features of the simulation) were combined with scattered electron flux distributions and tailored parameterizations of Cherenkov-light yield for each detector to estimate $A_{bias}$ and its uncertainty. The light yield varied strongly with the arrival angle and position of the electron on the detector. The light-yield parameterizations were developed to match the observed light-yield distribution by tuning optical parameters in a Geant4[59] optical photon transport simulation. The largest systematic uncertainty was associated with the optical modeling of the individual as-built detectors. This uncertainty was determined from the range in predicted $A_{bias}$ values obtained by varying optical parameters in the simulation while maintaining reasonable agreement with the measured light yield distributions. The predicted $A_{bias}$ values for each detector are shown in Extended Data Fig. 3c. The resulting averaged correction and its systematic uncertainty are $A_{bias} = 4.3 \pm 3.0$ ppb.

This result was consistent to within 1 ppb with a simpler, independent calculation based on a phenomenological approach, which made use of measured position distributions of the electron flux on the pre-radiators, and measured dependence of the light seen by each PMT on the track position on the detectors. Results obtained using measured distributions from the experiment's tracking system were compared with results obtained from simulation. In this approach, the effect of the position differences was scaled to match the observed left/right asymmetry. The sensitivity to various models for the position dependence of the asymmetries was found to be small.

**Determination of $A_{ep}$:** The measured asymmetry $A_{msr}$ was then corrected for incomplete beam polarization, the effects of various background processes, electromagnetic radiative corrections, and the finite acceptance of the detector, to form the fully-corrected electron-proton asymmetry $A_{ep}$, using

$$A_{ep} = R_{tot} \frac{\frac{A_{msr}}{P} - \sum_{i=1,3,4} f_i A_i}{1 - \sum_{i=1}^{4} f_i} \qquad (8)$$

where $R_{tot} = R_{RC} R_{Det} R_{Acc} R_{Q^2}$. The $f_i$ are dilutions to the signal and $A_i$ are false or background process asymmetries. The components of equation (8) are discussed below.

**$R_{RC}$:** The electromagnetic radiative correction factor $R_{RC} = 1.010 \pm 0.005$ accounts for the effect of internal and external bremsstrahlung of the incident electron, which can depolarize the electron and modify the momentum transfer $Q^2$ at the scattering vertex. $R_{RC}$ was determined using a Geant3[59] simulation by comparing results with and without bremsstrahlung enabled in the simulation.

**$R_{Det}$:** The Cherenkov detector analog response (i.e. the summed optical signal detected by the two photomultiplier tubes attached at each end of each detector) varied as a function of the arrival location on the detector of the scattered electron. The magnetic optics of the spectrometer also caused a correlation between the electron arrival location and $Q^2$, and therefore with the asymmetry. The correlation between the detector analog response and the $Q^2$ of each track was determined using the tracking system drift chambers, and the resulting correction to the measured asymmetry was $R_{Det} = 0.9895 \pm 0.0021$.

**$R_{Acc}$:** Due to the finite acceptance of the spectrometer, and the effect of radiative energy losses, $A_{msr}$ represents an average over a range of $Q^2$. Since the asymmetry varies strongly with $Q^2$, we used simulation to correct the averaged asymmetry $\langle A(Q^2) \rangle$ to the asymmetry that would arise from point scattering at the central $\langle Q^2 \rangle$, $\langle A(Q^2) \rangle$ using

$$R_{Acc} = \frac{A(\langle Q^2 \rangle)}{\langle A(Q^2) \rangle} = 0.977 \pm 0.002$$

**$R_{Q^2}$:** The central $\langle Q^2 \rangle$ for the experiment was determined from a Geant4 simulation that was benchmarked with measurements from the tracking system. The $\langle Q^2 \rangle$ was not identical for Run 1 and 2, due to minor differences in the beam energy, target location and spectrometer magnetic field, with Run 1 having a higher $\langle Q^2 \rangle$. The global fit of $A_{ep}$ vs. $Q^2$ (see Fig. 2) was used to determine the sensitivity of the asymmetry to small changes in $\langle Q^2 \rangle$. Run 2 was chosen as the reference for $\langle Q^2 \rangle$, and the Run 1 asymmetry was scaled from its measured $\langle Q^2 \rangle$ using $R_{Q^2} = 0.9928$ ($R_{Q^2} = 1$ for Run 2, by definition). The determination of the central $\langle Q^2 \rangle$ has a 0.45% relative uncertainty, dominated by the uncertainty on the collimator, target and main detector locations, and the beam energy determination. To simplify the global fitting, we decided to quote the $\langle Q^2 \rangle$ as exact and used the sensitivity $\partial A_{ep} / \partial Q^2$ to determine an effective error contribution to the asymmetry. This error on $R_{Q^2}$ was 0.0055 for both run periods. The acceptance-averaged $Q^2$, scattering angle and incident electron energy were $\langle Q^2 \rangle = 0.0248$ (GeV/c)$^2$, $\langle \theta \rangle = 7.90°$, and $\langle E_0 \rangle = 1.149$ GeV, respectively.

**$P$:** To achieve the goal of a reliable determination of the beam polarization ($P$) at <1% accuracy, two different techniques with precisely calculated analyzing powers were employed for redundancy. An existing Møller polarimeter[62] in experimental Hall C was used invasively 2 – 3 times per week. It measured the parity conserving cross-section asymmetry in the scattering of polarized beam electrons from polarized electrons in an iron foil target at low (typically ≲2 μA) beam currents. A newly installed, non-invasive Compton polarimeter[7] monitored the beam polarization continuously at the full production beam current of 180 μA. This device measured the parity-conserving asymmetry in the scattering of beam electrons from circularly polarized laser photons. For each run period, the averaged beam-polarization corrected asymmetry was computed in two ways – by correcting each ~6 minute period of data for the polarization measured during that interval, and by using an overall average beam polarization for the whole run period. The two methods gave the same result to a small fraction of the quoted uncertainty, so for simplicity the results using the overall average beam polarization are quoted here. The overall average beam polarizations for the two running periods were: $P_{\text{Run 1}} = (87.66 \pm 1.05)$ % and $P_{\text{Run 2}} = (88.71 \pm 0.55)$ %, where the uncertainties are predominantly systematic. For Run 1, the uncertainty was larger for two reasons: the Compton polarimeter was still being commissioned, so it was not used for this determination, and the Møller uncertainty was larger than usual due to the need to correct for the effects of an intermittent short circuit in one of the quadrupole magnets of the polarimeter. For Run 2, both polarimeters were fully functional and agreed well with each other, as shown in Extended Data Fig. 4. A dedicated direct comparison of the Møller and Compton polarimeters under identical beam conditions at low beam current was also performed. The two techniques agreed within the uncertainties for that measurement, $dP/P = 1\%$ and $dP/P = 0.73\%$ for the Compton and Møller, respectively[63].

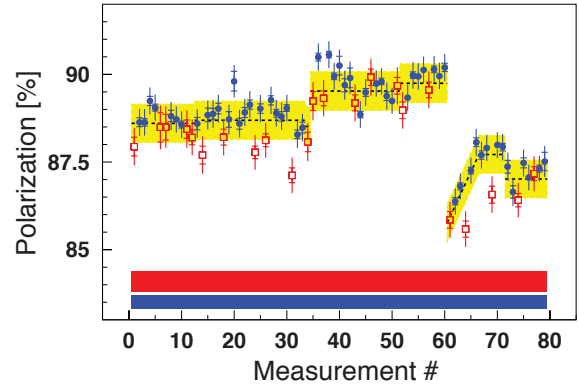

**Extended Data Figure 4 | Electron beam polarization.** Measurements from the Compton (closed blue circles) and Møller (open red squares) polarimeters during Run 2. Inner error bars denote statistical uncertainties and outer error bars show the statistical and point-to-point systematic uncertainties added in quadrature. Normalization, or scale-type, uncertainties are shown by the solid blue (Compton) and red (Møller) bands. All uncertainties are s.d. The yellow band shows the derived polarization values used in evaluation of the parity-violating asymmetry $A_{ep}$. The time dependence of the reported polarization is driven primarily by the continuous Compton measurements, with a small scale-correction (0.21%, not included in this figure) determined from an uncertainty-weighted global comparison of the Compton and Møller polarimeters.

**Physics Backgrounds (Target Windows).** Electrons scattered from the aluminum 7075 alloy entrance (0.11 mm thick) and exit (0.13 mm thick) windows of the hydrogen target caused the dominant background process ($f_i A_i = 37$ ppb for Run 1 and 38 ppb for Run 2). The parity-violating elastic asymmetry from aluminum is observed to be nearly an order of magnitude larger than that for the proton, due to the much larger weak charge of the aluminum nucleus[64], so even the small fraction of the detected yield arising from the windows required a significant correction to the measured asymmetry. The aluminum asymmetry was determined in dedicated data-taking with an aluminum target, made from the same block of material as the target windows, but with a thickness (3.7 mm) to match the radiation length of the hydrogen target. The range of scattering angles accepted from the upstream and downstream windows were different, which required a small kinematic correction to the measured alloy asymmetry to yield the asymmetry from the target windows $A_1 = 1.515 \pm 0.077$ ppm, see Extended Data Fig. 5. The uncertainty is dominated by statistics, but includes systematic uncertainties arising from the kinematic correction, among others. The fraction of the measured yield arising from the target windows $f_1 = (2.471 \pm 0.056)$ % (Run 1), $f_1 = (2.516 \pm 0.059)$ % (Run 2), was measured with low beam current on an evacuated target cell, and using simulation to correct for radiative effects due to the liquid hydrogen.

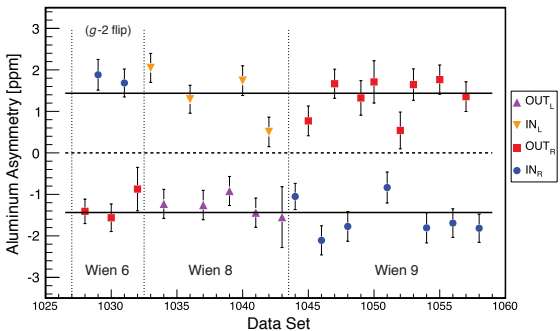

**Extended Data Figure 5 |Asymmetry from Aluminum.** Parity-violating asymmetry from the aluminum alloy target vs. dataset. All uncertainties are s.d. "IN" and "OUT" refer to the state of the insertable half-wave plate at the electron source, generating a 180° flip of the spin when IN. The subscripts denote the setting of the Wien filter as "L" or "R", corresponding to the presence or absence, respectively, of an additional 180° rotation of the spin direction of the electron beam. A period in which a further 180° flip was generated through $g - 2$ precession via a modified accelerator configuration is indicated. The combinations "OUT-R" and "IN-L" with no $g - 2$ spin flip expose the physical sign of the asymmetry. Solid lines represent the time-averaged values, and the horizontal dashed line indicates zero asymmetry. The vertical dashed lines delineate particular data subsets with a given Wien filter setting, labelled as Wien *i*.

**Physics Backgrounds (Beamline background dilution).** As described above (see $A_{BB}$), a component of the background came from scattering sources in the beamline. The dilution from this source was measured[58] to be $f_2 = (0.193$

$\pm 0.064)$ % by blocking two of the eight openings in the first collimator to eliminate the elastic electron signal from the target. The uncertainty accounts for variation in the value between detectors and under different beam conditions.

**Physics Backgrounds (Neutral background).** A possible contribution from low-energy neutral backgrounds arising from secondary interactions of the primary scattered electrons in the collimators and magnet structure was bounded[65] to be $f_3 < 0.30\%$ by subtracting $f_2$ from the total neutral background measured by the main detector after vetoing charged particles using thin scintillators. The asymmetry for this background was estimated[65] from a Geant4[59] simulation of the contributing processes to be $A_3 = -0.39 \pm 0.16$ ppm with the dominant contribution coming from secondary interactions of elastically scattered *e-p* electrons.

**Physics Backgrounds (Inelastic electrons).** An unavoidable background comes from inelastic electrons which have excited the target protons to the $\Delta(1232)$ resonance, a small fraction of which enter the acceptance of the spectrometer. The fraction of the yield from inelastic electrons was estimated using simulation to be $f_4 = (1.82 \pm 0.37) \times 10^{-4}$. To determine the correction to the asymmetry due to these events, we measured the parity-violating asymmetry in the $p \rightarrow \Delta(1232)$ region during a dedicated study with a reduced spectrometer current to concentrate these electrons on the detectors. Scaling this asymmetry up to the $Q^2$ of the elastic peak gives an inelastic asymmetry at the elastic peak $A_4 = -3.0 \pm 1.0$ ppm. Backgrounds from $\pi^-$s and other hadrons were negligible.

**Summary of Corrections to the Asymmetry and their Uncertainties**. The separate and combined measured asymmetries for Runs 1 and 2 are presented in Extended Data Table 2. Also shown is the breakdown of these uncertainties in terms of descending fractional significance. Extended Data Table 3 presents the numerical values of the "raw" asymmetry $A_{raw}$ along with all systematic and acceptance correction factors used to derive the observed measured asymmetry $A_{msr}$ and physics asymmetries $A_{ep}$ using equations (7) and (8). Correlated uncertainties, accounted for when the two runs were combined, are also listed.

| Period | Asymmetry (ppb) | Stat. Unc. (ppb) | Syst. Unc. (ppb) | Tot. Uncertainty (ppb) |
|---|---|---|---|---|
| Run 1 | -223.5 | 15.0 | 10.1 | 18.0 |
| Run 2 | -227.2 | 8.3 | 5.6 | 10.0 |
| Run 1 and 2 combined with correlations | -226.5 | 7.3 | 5.8 | 9.3 |

| Quantity | Run 1 error (ppb) | Run 1 fractional | Run 2 error (ppb) | Run 2 fractional |
|---|---|---|---|---|
| BCM Normalization: $A_{BCM}$ | 5.1 | 25% | 2.3 | 17% |
| Beamline Background: $A_{BB}$ | 5.1 | 25% | 1.2 | 5% |
| Beam Asymmetries: $A_{beam}$ | 4.7 | 22% | 1.2 | 5% |
| Rescattering bias: $A_{bias}$ | 3.4 | 11% | 3.4 | 37% |
| Beam Polarization: $P$ | 2.2 | 5% | 1.2 | 4% |
| Target windows: $A_{bl}$ | 1.9 | 4% | 1.9 | 12% |
| Kinematics: $R_{Q^2}$ | 1.2 | 2% | 1.3 | 5% |
| Total of others | 2.5 | 6% | 2.2 | 15% |
| Combined in quadrature | 10.1 | | 5.6 | |

**Extended Data Table 2 | Asymmetries and their corrections.** Top: Corrected asymmetries $A_{ep}$ for both data sets, and the combined value, with their statistical, systematic and total uncertainties (s.d.), in ppb. Bottom: Fractional quadrature contributions $(\sigma_i/\sigma_{tot})^2$ to the systematic uncertainty (s.d.) on $A_{ep}$ for Run 1 and 2. Only error sources with fractional contribution $\geq 5\%$ in one of the Runs are shown.

The fully corrected asymmetry for Run 1 is $A_{ep} = -223.5 \pm 15.0 \,(\text{stat}) \pm 10.1(\text{syst})$ ppb and for Run 2 is $A_{ep} = -227.2 \pm 8.3 \,(\text{stat}) \pm 5.6(\text{syst})$ ppb. The combined asymmetry is $A_{ep} = -226.5 \pm 7.3 \,(\text{stat}) \pm 5.8 \,(\text{syst})$ ppb.

**Data Quality.** Two representative tests of consistency and quality of the corrected asymmetries are presented below.

**Null Result:** The $Q_{weak}$ experiment employed signal phase locking on three independent techniques of polarized-beam helicity reversal to isolate the scattering asymmetry. These were the rapid (960 Hz) reversal, the insertion of a half-wave plate in the source laser optical path at 8-hour intervals, and the Wien-filter reversal at monthly intervals. The half-wave plate is a mechanical action and thus is unable to induce any false asymmetries electrically or magnetically. The Wien filter reversal provides a rejection of beam size (or focus) modulation-induced false asymmetry. By constructing an "out-of-phase" or "null" asymmetry $A_{null}$ from the latter two slow-reversal techniques, a determination can be made as to whether there are unaccounted for false asymmetries. The full data-set weighted null $A_{null} = -1.75 \pm 6.51$ ppb, which is consistent with zero, as expected.

| Quantity | Run 1 | Run 2 | Correlation |
|---|---|---|---|
| $A_{raw}$ | $-192.7 \pm 13.2$ ppb | $-170.7 \pm 7.3$ ppb | — |
| $A_T$ | $0 \pm 1.1$ ppb | $0 \pm 0.7$ ppb | 0 |
| $A_L$ | $1.3 \pm 1.0$ ppb | $1.2 \pm 0.9$ ppb | 1 |
| $A_{BCM}$ | $0 \pm 4.4$ ppb | $0 \pm 2.1$ ppb | 0.67 |
| $A_{BB}$ | $3.9 \pm 4.5$ ppb | $-2.4 \pm 1.1$ ppb | 0 |
| $A_{beam}$ | $18.5 \pm 4.1$ ppb | $0.0 \pm 1.1$ ppb | 0 |
| $A_{bias}$ | $4.3 \pm 3.0$ ppb | $4.3 \pm 3.0$ ppb | 1 |
| $A_{msr}$ | $-164.6 \pm 15.5$ ppb | $-167.5 \pm 8.4$ ppb | — |
| $P$ | $87.66 \pm 1.05$ % | $88.71 \pm 0.55$ % | 0.19 |
| $f_1$ | $2.471 \pm 0.056$ % | $2.516 \pm 0.059$ % | 1 |
| $A_1$ | $1.514 \pm 0.077$ ppm | $1.515 \pm 0.077$ ppm | 1 |
| $f_2$ | $0.193 \pm 0.064$ % | $0.193 \pm 0.064$ % | 1 |
| $f_3$ | $0.12 \pm 0.20$ % | $0.06 \pm 0.12$ % | 1 |
| $A_3$ | $-0.39 \pm 0.16$ ppm | $-0.39 \pm 0.16$ ppm | 1 |
| $f_4$ | $0.018 \pm 0.004$ % | $0.018 \pm 0.004$ % | 1 |
| $A_4$ | $-3.0 \pm 1.0$ ppm | $-3.0 \pm 1.0$ ppm | 1 |
| $R_{RC}$ | $1.010 \pm 0.005$ | $1.010 \pm 0.005$ | 1 |
| $R_{Det}$ | $0.9895 \pm 0.0021$ | $0.9895 \pm 0.0021$ | 1 |
| $R_{Acc}$ | $0.977 \pm 0.002$ | $0.977 \pm 0.002$ | 1 |
| $R_{Q^2}$ | $0.9928 \pm 0.0055$ | $1.0 \pm 0.0055$ | 1 |
| $R_{tot}$ | $0.9693 \pm 0.0080$ | $0.9764 \pm 0.0080$ | 1 |
| $\sum f_i$ | $2.80 \pm 0.22$ % | $2.78 \pm 0.15$ % | 1 |

**Extended Data Table 3 | Raw asymmetries and their corrections.** The raw measured asymmetries $A_{raw}$ for both run periods, and all the corrections for false asymmetries, backgrounds, beam polarization, detector acceptance, etc. applied to extract the final asymmetry $A_{ep}$ from $A_{raw}$ (see text). The $f_i$ are dilutions to the signal, $A_i$ are false or background process asymmetries, and $P$ and $R_i$ are multiplicative factors. The net multiplicative correction $R_{tot}$ and the total dilution are also indicated, as well as the values of $A_{msr}$, the asymmetry after the corrections for the false asymmetries (see equation 7). The correlations used to combine the two Runs are provided in the final column. Uncertainties are s.d.

**Asymmetry Measurements:** A plot of the observed main-detector asymmetry versus Wien-filter configuration is shown in Extended Data Fig. 6. Run 1 and Run 2 were separated by a six-month accelerator shutdown, during which numerous modifications were made to the experimental apparatus and accelerator. These included upgrading the electronics associated with the beam current measurement and increasing the number of associated BCMs. An electrically isolated helicity phase-locked beam-position stabilization system was enabled in the 6 MeV region of the injector during Run 2. This significantly improved the helicity-correlated stability of the beam delivered to the experiment. There were also radio-frequency associated electronics and superconducting cavity upgrades performed within the accelerator, unrelated to this experiment, as well as significant upgrades to both beam polarimeters. As a consequence, the contributions of many beam-related systematic effects meaningfully changed between the two run periods. However, the resulting fully corrected physics asymmetries agree well between the two run periods. This is evidence that, within the experiment's precision, the observed set of identified systematic effects is complete and their associated correction algorithms behave in a deterministic manner.

**Electroweak radiative corrections and extraction of $\sin^2 \theta_W$.** The weak mixing angle is obtained from the proton's weak charge, taking into account the energy-independent electroweak radiative corrections using[1]

$$4 \sin^2 \theta_W(0) =$$
$$1 - \frac{Q_W^p - \Box_{WW} - \Box_{ZZ} - \Box_{\gamma Z}(0)}{(\rho + \Delta_e)} + \Delta_e' . \quad (9)$$

Here $\Box_{\gamma Z}(0)$ refers to the remaining energy-independent piece of the $\gamma Z$ box diagram (the energy-dependent piece was discussed in Section B).

The accidental suppression of the proton's weak charge in the SM means that $Q_W^p$ is unusually sensitive to $\sin^2 \theta_W$. To see this quantitatively, our determination of $Q_W^p$ to 6.25% results in a $\sin^2 \theta_W$ precision of 0.46%. In contrast, the higher relative precision (0.59%) of the weak charge measurement[14,15] on $^{133}$Cs, dominated by the neutron's weak charge which is not suppressed in the SM, leads to a $\sin^2 \theta_W$ uncertainty of 0.81%, almost twice the uncertainty of our result.

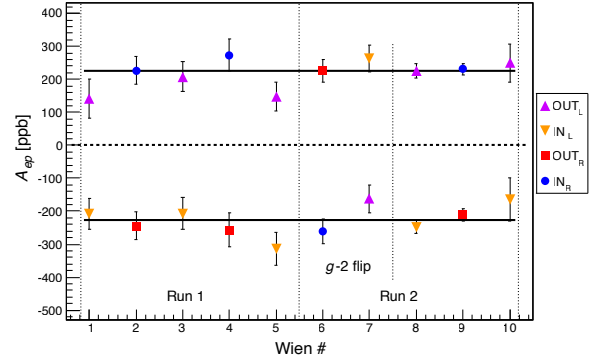

**Extended Data Figure 6 | Asymmetry from the proton.** Observed parity-violating asymmetry $A_{ep}$, after all corrections, vs. double Wien filter configuration. The Wien filter reversed the beam helicity at approximately monthly intervals. The subscripts denote the setting of the Wien filter as "L" or "R", corresponding to the presence or absence, respectively, of a 180° rotation of the spin direction of the electron beam. "IN" and "OUT" refer to the state of the insertable half-wave plate at the electron source, generating an additional 180° flip of the spin when IN. A period in which a further 180° flip was generated through $g - 2$ precession via a modified accelerator configuration is indicated. The combinations "OUT-R" and "IN-L" with no $g - 2$ flip expose the physical sign of the asymmetry. Solid lines represent the time-averaged values and the dashed line indicates zero asymmetry. The uncertainties (s.d.) shown are those of the corresponding $A_{\mathrm{msr}}$ values (see text) only, *i.e.* do not include time-independent uncertainties, so as to illustrate the time-stability of the results.

The radiative corrections appearing in equation (9) are described in ref. 1, but re-evaluated using the more recent input found in refs. 2 and 11. Although $Q_W^p$ in equation (6) is defined in the Thomson limit ($Q \ll m_e$) at $Q = 0$, it is determined from data in the scattering limit ($Q \gg m_e$). We chose to use radiative corrections in the Thomson limit from ref. 1 to calculate $\sin^2 \theta_W(0)$ from $Q_W^p$. Extended Data Table 4 lists these ingredients.

To compare our result for $\sin^2\theta_W(0)$, given in equation (9), with that of refs. 2 and 34, we added a correction $2\alpha/9\pi$, consistent with the definition of ref. 66, as discussed in ref. 20. The result is $\sin^2\theta_W(Q=0) = 0.2384 \pm 0.0011$ in the modified-minimal-subtraction ($\overline{MS}$) scheme[2]. Running it out to the $Q = 0.158$ GeV of the $Q_{\text{weak}}$ experiment (a correction of $-0.00012$), our $\sin^2\theta_W$ result is $\sin^2\theta_W(Q = 0.158\ \text{GeV}) = 0.2382 \pm 0.0111$.

| Term | Expression | Value | Reference |
|---|---|---|---|
| $\rho_{SC}$ | $1 + \Delta_\rho$ | 1.00066 | 1, 2 |
| $\Delta_e$ | $-\alpha/2\pi$ | $-0.001161$ | 1, 2 |
| $\Delta'_e$ | $-\frac{\alpha}{3\pi}(1 - 4\hat{s}^2)\left[\ln\left(\frac{M_Z^2}{m_e^2}\right) + \frac{1}{6}\right]$ | $-0.001411$ | 1, 2 |
| $\hat{\alpha}$ | $\equiv \alpha(M_Z)$ | $1/127.95$ | 1, 2 |
| $\hat{s}^2$ | $= 1 - \hat{c}^2 \equiv \sin^2\theta_W(M_Z)$ | 0.23129 | 1, 2 |
| $\alpha_s(M_Z^2)$ | | 0.12072 | 67 |
| $\square_{WW}$ | $\frac{\hat{\alpha}}{4\hat{s}^2}\left[2 + 5\left(1 - \frac{\alpha_s(M_Z^2)}{\pi}\right)\right]$ | 0.01831 | 1, 2 |
| $\square_{ZZ}$ | $\frac{\hat{\alpha}}{4\hat{s}^2\hat{c}^2}\left[9/4 - 5\hat{s}^2\right](1 - 4\hat{s}^2 + 8\hat{s}^4)\left(1 - \frac{\alpha_s(M_Z^2)}{\pi}\right)$ | 0.00185 | 1, 2 |
| $\square_{\gamma Z}$ | axial-vector hadron piece of $\square_{\gamma Z}$: $\Re e\, \square_{\gamma Z}^A$ | 0.0044 | 11 |

**Extended Data Table 4 | Radiative corrections.** Numerical values used for the electroweak radiative corrections in equation 9.

**Data Availability Statement:** The 200 TB of raw data are stored at the Jefferson Lab data silo. Derived data supporting the findings of this study are available from the corresponding author upon request.

**Code Availability Statement:** The software utilized for data management and analysis consisted of commercial and publicly available codes, plus experiment-specific software. Jefferson Lab's data management plan is available at:

https://scicomp.jlab.org/DataManagementPlan.pdf. The experiment-specific software is stored in a version management system (SVN & GIT) and archived at the data storage facilities of Jefferson Lab in accordance with existing US regulations. Requests for this material should be addressed to the corresponding authors of this paper.